\documentstyle[12pt]{article}
\begin{document}

\begin{titlepage}

\null

\begin{flushright}
 KOBE-TH-93-11 \\
 December 1993
\end{flushright}

\vspace{1cm}
\begin{center}
 {\Large\bf
 Fermion Currents on Asymmetric Orbifolds
 \par}
\vspace{2.5cm}
\baselineskip=7mm
 {\large
  Toshihiro Sasada
  \par}
\vspace{5mm}
{\sl
  Department of Physics, Kobe University\\
  Rokkodai, Nada, Kobe 657, Japan
  \par}

\vspace{3cm}
 {\large\bf Abstract}
\end{center}
\par

We study whether orbifold models are equivalently rewritten into
torus models in the case of fermionic string theories.
It is pointed out that symmetric orbifold models cannot be
rewritten into torus models in the case of fermionic string
theories because of the absence of twist-untwist intertwining
currents on the orbifold models.
We present a list of current algebras on asymmetric $Z_N$-orbifold
models of type II superstring theories with inner automorphisms of
Lie algebra lattices of the $A_n$ series.
It turns out that whether an asymmetric orbifold model is rewritten
into a torus model depends on the specific choice of a momentum
lattice and an inner automorphism of the lattice.

\end{titlepage}

\baselineskip=7mm


Various methods have been used to construct
four-dimensional string models.
However, the relationship between different methods has not been
so fully understood.
One of the examples showing such relationship between different
constructions is the torus-orbifold equivalence
\cite{FLM}--\cite{ISST}
in bosonic string theories.
If the $Z_N$-transformation of an orbifold model
is an inner automorphism of the momentum lattice,
then the $Z_N$-transformation is equivalent to a shift on the lattice.
The orbifold model associated with a shift is equivalent to
a torus model in the case of
bosonic string theories\footnote{Examples which suggest strongly
the torus-orbifold equivalence with outer automorphisms
of the momentum lattices are discussed in ref. \cite{ISST}
in the case of bosonic string
theories.}.
In this paper, we shall investigate whether orbifold models are
equivalently rewritten into torus models in the case of fermionic
(i.e. heterotic or type II) string theories, which has not
been studied in detail before.

We shall define the fermion currents on an orbifold model as the
currents of the Ka\v{c}-Moody algebra generated by the NSR fermions
on the orbifold model.
Such fermion currents will generate a Ka\v{c}-Moody algebra
which contain the $SO(2)$ Ka\v{c}-Moody algebra generated by the
transverse space-time NSR fermions in
the light-cone gauge.
On the other hand, fermions on a torus model are all NSR fermions
and will generate
an $SO(8)$ Ka\v{c}-Moody algebra in the light-cone gauge.
Thus, the necessary condition for the orbifold model to be rewritten
into a torus model is that the fermion currents on the orbifold
model should generate an $SO(8)$ Ka\v{c}-Moody algebra
and
it turns out that
this is also the sufficient condition for the orbifold
model with an inner automorphism of the momentum lattice
to be rewritten into a torus model.

It should be noted that, in order for the fermion currents on an
orbifold model to generate an $SO(8)$ Ka\v{c}-Moody
algebra, it is necessary to exist
the twist-untwist intertwining currents \cite{ISST}--\cite{DGM}
which convert untwisted string states to twisted ones in the
orbifold model.
The reason for this is that, in each twisted or untwisted
sector, the unbroken Ka\v{c}-Moody algebra generated by the fermions
on the orbifold model
is always smaller
than the $SO(8)$ algebra generated by the fermions on a torus
model.
Therefore, symmetric orbifold models cannot be rewritten into
torus models in the
case of fermionic string theories since there is no twist-untwist
intertwining current on symmetric orbifold models.
In the following, we will construct asymmetric $Z_N$-orbifold
models of type II superstring theories
with inner automorphisms of Lie
algebra lattices of the $A_n$ series and investigate whether such
asymmetric orbifold models could equivalently be rewritten into
torus models.

In the construction of an orbifold model, we start with a
6-dimensional torus compactification of type II superstring
theories
which is specified by a $(6+6)$-dimensional even self-dual
lattice $\Gamma^{6,6}$ \cite{NSW}.
The left- and right-moving momentum
$(p_L^i, p_R^i)$ $ (i=1, \ldots ,6)$ lies on the lattice
$\Gamma^{6,6}$.
Let $g$ be a group element which generates a cyclic group
$Z_N$.
The $g$ is defined to act on the left- and right-moving string
coordinate $(X_L^i, X_R^i)$ $ (i=1,\ldots,6)$ by
\begin{equation}
g: (X_L^i,X_R^i) \rightarrow (U_L^{ij} X_L^j,  U_R^{ij} X_R^j),
\label{eqn:ZNtr}
\end{equation}
where $U_L$ and $U_R$ are rotation matrices which satisfy
$U_L^N=U_R^N=1$.
The $g$ acts on the left-movers and the right-movers
differently.
The $Z_N$-transformation must be an automorphism of the
lattice $\Gamma^{6,6}$, i.e.,
\begin{equation}
(U_L^{ij} p_L^j, U_R^{ij} p_R^j) \in \Gamma^{6,6}
\ \  {\rm for \ all} \ (p_L^i,p_R^i) \in \Gamma^{6,6}.
\end{equation}
The action of the operator $g$ on the left- and right-moving
fermions on the orbifold is given by $U_L$ and $U_R$
rotations, respectively.

Let us consider the $g^{\ell}$-twisted sector in which
strings close up to the $g^{\ell}$-action.
We denote the eigenvalues of $U_L^{\ell}$ and $U_R^{\ell}$
by $ \{ e^{i 2\pi r_L^a},e^{-i 2\pi r_L^a} ; a=1,2,3 \} $ and
$ \{ e^{i 2\pi r_R^a},e^{-i 2\pi r_R^a} ; a=1,2,3 \} $,
respectively.
Let $N_{\ell}$ be the minimum positive integer such that
$(g^{\ell})^{N_{\ell}} = 1$.
The necessary conditions for one-loop modular
invariance are for $N_{\ell}$ even
\begin{equation}
N_\ell \sum_{a=1}^3 r_L^a = 0 \ \  \bmod 2,
\label{eqn:mod1}
\end{equation}
\begin{equation}
N_\ell \sum_{a=1}^3 r_R^a = 0 \ \  \bmod 2,
\label{eqn:mod2}
\end{equation}
\begin{equation}
  p_L^i (U_L^{\frac{N_\ell}{2}})^{ij} p_L^j
- p_R^i (U_R^{\frac{N_\ell}{2}})^{ij} p_R^j
= 0 \ \  \bmod 2
\label{eqn:mod3}
\end{equation}
for all $(p_L^i,p_R^i) \in \Gamma^{6,6}$;
for $N_\ell$ odd, there is no condition for one-loop
modular invariance
\cite{NSV1}.
These are called the left-right level matching
conditions and it has been proved that these
are also sufficient conditions for one-loop modular invariance
\cite{V,NSV2}.

Suppose that the $Z_N$-transformation is an inner automorphism of
the momentum lattice.
Then, it can be shown that the $Z_N$-transformation is equivalent
to a shift as follows \cite{S}:
Let us consider the $g^{\ell}$-sector (the untwisted sector for
$ \ell = 0 $ and the twisted sector for $ \ell = 1,\ldots,N-1 $).
Since the rank of the unbroken current algebra in each sector
is equal to the dimension of the orbifold, we can always construct
the $Z_N$-invariant operators $P_L^{\prime i}(z)$ and
$P_R^{\prime i}(\bar{z})$ $(i = 1,\ldots,6)$ such that
\begin{equation}
g (P_L^{\prime i}(z), P_R^{\prime i}(\bar{z})) g^{-1}
= (P_L^{\prime i}(z), P_R^{\prime i}(\bar{z})) ,
\label{eqn:inv}
\end{equation}
and
\begin{equation}
  P_L^{\prime i}(w) P_L^{\prime j}(z)
= {\delta^{ij}\over (w-z)^2} + ({\rm regular \ terms}),
\label{eqn:opeL}
\end{equation}
\begin{equation}
  P_R^{\prime i}(\bar{w})P_R^{\prime j}(\bar{z})
= {\delta^{ij}\over (\bar{w}-\bar{z})^2} + ({\rm regular\ terms}),
\label{eqn:opeR}
\end{equation}
where $g$ is the operator which generates the $Z_N$-transformation
in the $g^{\ell}$-sector.
It follows from (\ref{eqn:opeL}) and (\ref{eqn:opeR}) that
$P_L^{\prime i}(z)$ and $P_R^{\prime i}(\bar{z})$ can be expanded
as
\begin{equation}
P_L^{\prime i}(z) \equiv
i \partial_z X_L^{\prime i}(z) \equiv
\sum_{n\in Z}\alpha_{L n}^{\prime i} z^{-n-1},
\label{eqn:modeL}
\end{equation}
\begin{equation}
P_R^{\prime i}(\bar{z}) \equiv
i \partial_{\bar{z}} X_R^{\prime i}(\bar{z}) \equiv
\sum_{n\in Z}\alpha_{R n}^{\prime i} {\bar{z}}^{-n-1} ,
\label{eqn:modeR}
\end{equation}
with
\begin{equation}
[ \alpha_{Lm}^{\prime i}, \alpha_{Ln}^{\prime j}]
= m \delta^{ij} \delta_{m+n,0},
\end{equation}
\begin{equation}
[ \alpha_{Rm}^{\prime i}, \alpha_{Rn}^{\prime j} ]
= m\delta^{ij}\delta_{m+n,0},
\end{equation}
where $\alpha_{L0}^{\prime i} = p_L^{\prime i}$ and
$\alpha_{R0}^{\prime i} = p_R^{\prime i}$.
Since $P_L^{\prime i}(z)$ and $P_R^{\prime i}(\bar{z})$
are invariant under the $Z_N$-transformation, the string coordinate
in the new basis transforms as
\begin{equation}
g (X_L^{\prime i}(z), X_R^{\prime i}(\bar{z})) g^{-1}
= (X_L^{\prime i}(z)+2\pi v_L^i, X_R^{\prime i}(\bar{z})-2\pi v_R^i),
\end{equation}
for some constant vector $(v_L^i, v_R^i)$.
This implies that the string coordinate
$( X_L^{\prime i}(z), X_R^{\prime i}(\bar{z}) )$ in the
$g^{\ell}$-sector obeys the following boundary condition:
\begin{equation}
(X_L^{\prime i}(e^{2\pi i}z), X_R^{\prime i}(e^{-2\pi i}\bar{z}))
= (X_L^{\prime i}(z)+2\pi \ell v_L^i,
   X_R^{\prime i}(\bar{z})-2\pi \ell v_R^i)
+ ({\rm torus \  shift}),
\end{equation}
and hence that the eigenvalues of the momentum
$(p_L^{\prime i}, p_R^{\prime i})$ in the new basis are of the form
\begin{equation}
(p_L^{\prime i},p_R^{\prime i}) \in \Gamma^{6,6} + \ell (v_L^i, v_R^i).
\end{equation}
Since the lattice $\Gamma^{6,6}$ is self-dual, the shift vector
$(v_L^i, v_R^i)$ must satisfy
\begin{equation}
N (v_L^i,v_R^i) \in \Gamma^{6,6} ,
\end{equation}
\begin{equation}
\ell (v_L^i,v_R^i) \not \in \Gamma^{6,6}\  (\ell = 1, \ldots ,N-1).
\end{equation}

We may bosonize the fermions on the orbifold.
The GSO projected left- and right-moving NSR fermions are represented
by the bosons $\phi_L^t(z)$ and $\phi_R^t(\bar{z})$ $(t = 1,\ldots,4)$,
respectively. The momentum $(p_L^t, p_R^t)$ of the bosons each
lies on the weight lattice of $SO(8)$.
The momentum in the vector conjugacy class corresponds to the state
in the NS sector.
The momentum in the spinor or conjugate spinor conjugacy class
corresponds to the state in the R sector.
We denote the eigenvalues of $U_L$ and $U_R$ as
$\{ e^{i 2\pi \zeta_L^a}, e^{-i 2\pi \zeta_L^a} ; a=1,2,3 \}$ and
$\{ e^{i 2\pi \zeta_R^a}, e^{-i 2\pi \zeta_R^a} ; a=1,2,3 \}$,
respectively.
The $Z_N$-transformation acts on $\phi_L^t(z)$ and $\phi_R^t(\bar{z})$
as a shift:
\begin{equation}
g (\phi_L^t(z), \phi_R^t(\bar{z})) g^{-1} =
   (\phi_L^t(z) + 2\pi v_L^t, \phi_R^t(\bar{z}) - 2\pi v_R^t),
\end{equation}
where the shift vector $(v_L^t, v_R^t)$ are given by
\begin{equation}
v_L^t = (0, \zeta_L^a ),
\end{equation}
\begin{equation}
v_R^t = (0, \zeta_R^a ).
\end{equation}
Thereby, the eigenvalues of the momentum $(p_L^t, p_R^t)$ in the
$g^{\ell}$-sector are shifted from the ones in the untwisted sector
by the constant vector $\ell (v_L^t, v_R^t)$.

In this bosonized form, the operator $g$ will be given by
\begin{equation}
g = \eta_{(\ell)} \exp [ i 2 \pi
             ( p_L^{\prime i} v_L^i - p_R^{\prime i} v_R^i
                    +p_L^t v_L^t - p_R^t v_R^t )  ],
\end{equation}
where $\eta_{(\ell)}$ is a constant phase with $(\eta_{(\ell)})^N=1$.
The phase $\eta_{(\ell)}$ is determined in exactly the same way as
in ref. \cite{ISST} by one-loop modular invariance:
\begin{equation}
  \eta_{(\ell)} = \exp [ - i \pi \ell (  ( v_L^i )^2 - ( v_R^i )^2
                                     + ( v_L^t )^2 - ( v_R^t )^2 )  ].
\end{equation}
Every physical state must obey the condition $g=1$ because it must
be invariant under the $Z_N$-transformation.
Thus, the allowed momentum eigenvalues
$(p_L^{\prime i}, p_R^{\prime i}; p_L^t, p_R^t)$ of the physical
states in the $g^{\ell}$-sector must satisfy the following condition:
\begin{equation}
  p_L^{\prime i} v_L^i - p_R^{\prime i} v_R^i
+ p_L^t v_L^t - p_R^t v_R^t
-\frac{1}{2} \ell ( ( v_L^i )^2 - ( v_R^i )^2
          + ( v_L^t )^2 - ( v_R^t )^2 )
= 0 \ \ \bmod 1.
\end{equation}

We may now discuss the asymmetric $Z_N$-orbifold models
with inner automorphisms of Lie algebra lattices of
the $A_n$ series.
We take the lattice $\Gamma^{6,6}$ of an
asymmetric $Z_N$-orbifold model to be of the form:
\begin{equation}
\Gamma^{6,6}=\{ (p_L^i,p_R^i) \vert  p^i_L,p^i_R\in \Lambda^{\ast},
p_L^i-p_R^i \in \Lambda  \},
\label{eqn:ENlat}
\end{equation}
where $\Lambda$ is a 6-dimensional lattice and
$\Lambda^{\ast}$ is the dual lattice of $\Lambda$.
It turns out that $\Gamma^{6,6}$ is Lorentzian even self-dual if
$\Lambda$ is even integral.
In the following, we will take $\Lambda$ in eq. (\ref{eqn:ENlat})
to be the products of root lattices of $A_n$ algebras
with the squared length of the root vectors normalized to
two \cite{EN}.
The left- and right-rotation matrices of the
$Z_N$-transformation in eq. (\ref{eqn:ZNtr}) are taken from
the Weyl group elements of the root lattices of the
$A_n$ algebras.
Then it is easy to see that such a $Z_N$-transformation is an
automorphism of the lattice $\Gamma^{6,6}$ in eq. (\ref{eqn:ENlat}).

The conjugacy classes of the Weyl groups of all simple Lie algebras
have been classified in ref. \cite{C}.
In the following, we will give the results concerning the conjugacy
classes of the Weyl groups of Lie algebras of the $A_n$ series.
We may use the orthonormal basis $\{ e_i; i=1,\ldots,n+1\}$
to describe the roots of the $A_n$ algebra.
Then the roots of the $A_n$ algebra is given by
$\{ \pm(e_i - e_j) \}$.
In this basis, any element of the Weyl group of the $A_n$ algebra
is given by  a permutation of the $n+1$ basis vectors.
This permutation can be expressed as a product of the disjoint
cyclic permutations of the basis vectors.
Let $[m_1,m_2,\ldots]$ be the structure of such cyclic permutations.
Then there is a one-to-one correspondence between the structures
of the cyclic permutations and the conjugacy classes of the
Weyl group elements.
Hereafter we shall call $[m_1,m_2,\ldots]$ the cycle-type of
the automorphism of the Lie algebra lattice of the $A_n$ series.

Although the left- and right-rotations of the $Z_N$-transformations
in eq. (\ref{eqn:ZNtr}) can be defined for arbitrary elements of
the Weyl groups, the conditions
(\ref{eqn:mod1}), (\ref{eqn:mod2}) and (\ref{eqn:mod3}) for the
modular invariance put severe restrictions on the allowed
left- and right-rotations of the $Z_N$-transformations.
All the models we have to consider are shown in table 1.
The root lattice $\Lambda$ associated with the momentum lattice
$\Gamma^{6,6}$ in eq. (\ref{eqn:ENlat}) is given
in the second column of table 1.
The left- and right-moving cycle-types $C_L$ and $C_R$ of
the automorphism of the momentum lattice are given
in the third and fourth columns of table 1, respectively.

Now we will investigate fermion currents on the asymmetric
$Z_N$-orbifold models listed in table 1
in order to study whether such
orbifold models are equivalently rewritten into
torus models in the case of fermionic string theories.
However, it is not easy in general to examine current
algebras on fermionic string theories due to
the (generalized) GSO projections.
Then we may use the bosonic string map \cite{CENST}
and investigate current algebras on the corresponding
bosonic string models.
In the case of four-dimensional string models,
the bosonic string map works as follows:
The light-cone $SO(2)$ Ka\v{c}-Moody algebra generated by
the NSR fermions in the fermionic
string theory is replaced by an $SO(10) \times E_8$ Ka\v{c}-Moody
algebra in the bosonic string theory.
This is done in such a way that the $SO(2)$ Ka\v{c}-Moody characters
of the fermionic string theory are mapped to the
$SO(10) \times E_8$ Ka\v{c}-Moody characters of the bosonic string
theory preserving the modular transformation properties of the
Ka\v{c}-Moody characters.
Under this map, the gravitino state
(i.e. the space-time supercharges) in the fermionic string
theory becomes a set of operators of conformal weight one
transforming as a spinor of the $SO(10)$ in the bosonic
string theory.
Then it can be shown that, in the bosonic string theory,
these operators of weight one should extend the
$SO(10)$ Ka\v{c}-Moody algebra to one of the exceptional algebras
of $E_6$, $E_7$ and $E_8$ \cite{LSW}.
On the corresponding bosonic string models, we can easily
see that the orbifold models must be asymmetric in order to possess
twist-untwist intertwining currents.
The reason for this is that the left- and right-conformal weights,
$h$ and $\bar{h}$, of the ground states of any twisted sector are
both positive and equal in the case of the symmetric orbifold
models.

Using the bosonic string map, we investigate the fermion
currents on the asymmetric orbifold models listed in table 1.
It should be noted that we must investigate the full current
algebras on the orbifold models which possess twist-untwist
intertwining currents in order to determine fermion currents on
the orbifold models.
The results are summarized in table 2.
The fermion currents $F_L$ and $F_R$ in the left- and right-moving
degrees of freedom are given in the first and second columns of
table 2, respectively.
The numbers of space-time supercharges $N_L$ and $N_R$ from the
left- and right-moving degrees of freedom are given in the third and
fourth columns of table 2, respectively.
The left- and right-moving remaining current algebras
$G_L$ and $G_R$ of the asymmetric orbifold models are given in the
fifth and sixth columns of table 2, respectively.
The model numbers of the corresponding asymmetric orbifold models
are given in the last column of table 2.
In order for an asymmetric orbifold model
to be rewritten into a torus model,
it is necessary that both the left- and
right-moving fermion currents $F_L$ and $F_R$
should generate the $SO(8)$ Ka\v{c}-Moody algebras.
In fact, this is also the sufficient condition
for the asymmetric orbifold model with the
inner automorphism of the momentum
lattice to be rewritten into a torus model
since the remaining current algebras
$G_L$ and $G_R$ can be constructed in terms
of the string coordinate on the 6-dimensional
torus whose momentum lies on the lattice
determined by the physical momentum of the
asymmetric orbifold models.

We have discussed whether orbifold models are equivalently
rewritten into torus models in the case of fermionic string theories
and presented a list of fermion currents on asymmetric $Z_N$-orbifold
models of type II superstring theories
with inner automophisms of Lie algebra lattices
of the $A_n$ series.
We have found that some of the asymmetric orbifold models can be
rewritten into torus models owing to the existence of the
twist-untwist intertwining currents and that
whether an asymmetric orbifold model
is rewritten into a torus model depends on the specific choice of the
momentum lattice and an inner automorphism of the lattice.
These results in the fermionic string theories
differ entirely from the ones in the
bosonic string theories because all the
symmetric and asymmetric bosonic orbifold models
with inner automorphisms of momentum lattices
can equivalently be rewritten
into the torus models.

It will be straightforward to apply our analysis to heterotic
string theories.
On the other hand, it would be of interest to classify all asymmetric
$Z_N$-orbifold models of type II superstring theories
with inner automorphisms
of the momentum lattices.
Work in this direction is in progress and will be reported elsewhere.

\bigskip

The author would like to thank Dr. M. Sakamoto for reading the
manuscript and useful comments.


\newpage



\newpage

\pagestyle{empty}


\begin{table}

\caption{
Asymmetric $Z_N$-orbifold models with inner automorphisms of Lie
algebra lattices of the $A_n$ series.
$\Lambda$ denotes the root lattice associated with the momentum
lattice.
$C_L$ and $C_R$ denote the left- and right-moving cycle-types of the
automorphism of the momentum lattice, respectively, where semicolons
separate the direct products of the lattices.
}

\vspace{3 mm}

\centering

\begin{tabular}{|c|c|c|c|}
\hline
No. &$ \Lambda
   $&$ C_L
   $&$ C_R
   $\\
\hline
1   &$  A_2 \times A_1 \times A_1 \times A_1 \times A_1
   $&$ [1^3;1^2;1^2;1^2;1^2]
   $&$ [3;1^2;1^2;1^2;1^2]
   $\\
2   &$  A_2 \times A_2 \times A_1 \times A_1
   $&$ [1^3;1^3;1^2;1^2]
   $&$ [1^3;3;1^2;1^2]
   $\\
3   &$   A_2 \times A_2 \times A_1 \times A_1
   $&$ [1^3;1^3;1^2;1^2]
   $&$ [3;3;1^2;1^2]
   $\\
4   &$   A_2 \times A_2 \times A_1 \times A_1
   $&$ [1^3;3;1^2;1^2]
   $&$ [3;1^3;1^2;1^2]
   $\\
5   &$   A_2 \times A_2 \times A_1 \times A_1
   $&$ [1^3;3;1^2;1^2]
   $&$ [3;3;1^2;1^2]
   $\\
6   &$  A_2 \times A_2 \times A_2
   $&$ [1^3;1^3;1^3]
   $&$ [1^3;1^3;3]
   $\\
7   &$  A_2 \times A_2 \times A_2
   $&$ [1^3;1^3;1^3]
   $&$ [1^3;3;3]
   $\\
8   &$  A_2 \times A_2 \times A_2
   $&$ [1^3;1^3;3]
   $&$ [1^3;3;1^3]
   $\\
9   &$  A_2 \times A_2 \times A_2
   $&$ [1^3;1^3;3]
   $&$ [1^3;3;3]
   $\\
10  &$  A_2 \times A_2 \times A_2
   $&$ [1^3;1^3;1^3]
   $&$ [3;3;3]
   $\\
11  &$  A_2 \times A_2 \times A_2
   $&$ [1^3;1^3;3]
   $&$ [3;3;1^3]
   $\\
12  &$  A_2 \times A_2 \times A_2
   $&$ [1^3;1^3;3]
   $&$ [3;3;3]
   $\\
13  &$  A_2 \times A_2 \times A_2
   $&$ [1^3;3;3]
   $&$ [3;1^3;3]
   $\\
14  &$  A_2 \times A_2 \times A_2
   $&$ [1^3;3;3]
   $&$ [3;3;3]
   $\\
15  &$  A_3 \times A_1 \times A_1 \times A_1
   $&$ [1^4;1^2;1^2;1^2]
   $&$ [3,1;1^2;1^2;1^2]
   $\\
16  &$  A_3 \times A_2 \times A_1
   $&$ [1^4;1^3;1^2]
   $&$ [1^4;3;1^2]
   $\\
17  &$  A_3 \times A_2 \times A_1
   $&$ [1^4;1^3;1^2]
   $&$ [3,1;1^3;1^2]
   $\\
18  &$  A_3 \times A_2 \times A_1
   $&$ [1^4;1^3;1^2]
   $&$ [3,1;3;1^2]
   $\\
19  &$  A_3 \times A_2 \times A_1
   $&$ [1^4;3;1^2]
   $&$ [3,1;1^3;1^2]
   $\\
20  &$  A_3 \times A_2 \times A_1
   $&$ [1^4;3;1^2]
   $&$ [3,1;3;1^2]
   $\\
21  &$  A_3 \times A_2 \times A_1
   $&$ [3,1;1^3;1^2]
   $&$ [3,1;3;1^2]
   $\\
22  &$  A_3 \times A_3
   $&$ [1^4;1^4]
   $&$ [1^4;3,1]
   $\\
23  &$  A_3 \times A_3
   $&$ [1^4;1^4]
   $&$ [3,1;3,1]
   $\\
24  &$  A_3 \times A_3
   $&$ [1^4;3,1]
   $&$ [3,1;1^4]
   $\\
25  &$  A_3 \times A_3
   $&$ [1^4;3,1]
   $&$ [3,1;3,1]
   $\\
26  &$  A_4 \times A_1 \times A_1
   $&$  [1^5;1^2;1^2]
   $&$ [3,1^2;1^2;1^2]
   $\\
27  &$  A_4 \times A_1 \times A_1
   $&$  [1^5;1^2;1^2]
   $&$ [5;1^2;1^2]
   $\\
28  &$  A_4 \times A_1 \times A_1
   $&$  [3,1^2;1^2;1^2]
   $&$ [5;1^2;1^2]
   $\\
\hline
\end{tabular}

\end{table}


\clearpage


\setcounter{table}{0}

\begin{table}

\caption{
(continued)
}

\vspace{3 mm}

\centering

\begin{tabular}{|c|c|c|c|}
\hline
No. &$ \Lambda
   $&$ C_L
   $&$ C_R
   $\\
\hline
29  &$  A_4 \times A_2
   $&$  [1^5;1^3]
   $&$ [1^5;3]
   $\\
30  &$  A_4 \times A_2
   $&$  [1^5;1^3]
   $&$ [3,1^2;1^3]
   $\\
31  &$  A_4 \times A_2
   $&$  [1^5;1^3]
   $&$ [3,1^2;3]
   $\\
32  &$  A_4 \times A_2
   $&$  [1^5;3]
   $&$ [3,1^2;1^3]
   $\\
33  &$  A_4 \times A_2
   $&$  [1^5;3]
   $&$ [3,1^2;3]
   $\\
34  &$  A_4 \times A_2
   $&$  [1^5;1^3]
   $&$ [5;1^3]
   $\\
35  &$  A_4 \times A_2
   $&$  [1^5;1^3]
   $&$ [5;3]
   $\\
36  &$  A_4 \times A_2
   $&$  [1^5;3]
   $&$ [5;1^3]
   $\\
37  &$  A_4 \times A_2
   $&$  [1^5;3]
   $&$ [5;3]
   $\\
38  &$  A_4 \times A_2
   $&$  [3,1^2;1^3]
   $&$ [3,1^2;3]
   $\\
39  &$  A_4 \times A_2
   $&$  [3,1^2;1^3]
   $&$ [5;1^3]
   $\\
40  &$  A_4 \times A_2
   $&$  [3,1^2;1^3]
   $&$ [5;3]
   $\\
41  &$  A_4 \times A_2
   $&$  [3,1^2;3]
   $&$ [5;1^3]
   $\\
42  &$  A_4 \times A_2
   $&$  [3,1^2;3]
   $&$ [5;3]
   $\\
43  &$  A_4 \times A_2
   $&$  [5;1^3]
   $&$ [5;3]
   $\\
44  &$  A_5 \times A_1
   $&$  [1^6;1^2]
   $&$ [3,1^3;1^2]
   $\\
45  &$  A_5 \times A_1
   $&$  [1^6;1^2]
   $&$ [3,3;1^2]
   $\\
46  &$  A_5 \times A_1
   $&$  [1^6;1^2]
   $&$ [5,1;1^2]
   $\\
47  &$  A_5 \times A_1
   $&$  [3,1^3;1^2]
   $&$ [3,3;1^2]
   $\\
48  &$  A_5 \times A_1
   $&$  [3,1^3;1^2]
   $&$ [5,1;1^2]
   $\\
49  &$  A_5 \times A_1
   $&$  [3,3;1^2]
   $&$ [5,1;1^2]
   $\\
50  &$  A_6
   $&$  [1^7]
   $&$ [3,1^4]
   $\\
51  &$  A_6
   $&$  [1^7]
   $&$ [3,3,1]
   $\\
52  &$  A_6
   $&$  [1^7]
   $&$ [5,1^2]
   $\\
53  &$  A_6
   $&$  [1^7]
   $&$ [7]
   $\\
54  &$  A_6
   $&$  [3,1^4]
   $&$ [3,3,1]
   $\\
55  &$  A_6
   $&$  [3,1^4]
   $&$ [5,1^2]
   $\\
56  &$  A_6
   $&$  [3,1^4]
   $&$ [7]
   $\\
57  &$  A_6
   $&$  [3,3,1]
   $&$ [5,1^2]
   $\\
58  &$  A_6
   $&$  [3,3,1]
   $&$ [7]
   $\\
59  &$  A_6
   $&$  [5,1^2]
   $&$ [7]
   $\\
\hline
\end{tabular}

\end{table}


\clearpage


\begin{table}

\caption{
Current algebras on asymmetric $Z_N$-orbifold models. $F_L$ and $F_R$
denote the left- and right-moving fermion currents on asymmetric
orbifold models, respectively.
$N_L$ and $N_R$ denote the numbers of space-time supercharges from
left- and right-moving degrees of freedom, respectively.
$G_L$ and $G_R$ denote the left- and right-moving remaining current
algebras on the asymmetric orbifold models, respectively.
In the last column, the model numbers of the corresponding asymmetric
orbifold models are indicated.
}

\vspace{3 mm}

\centering

\begin{tabular}{|c|c|c|c|c|c|c|}
\hline
$
   F_L    $&$
   F_R    $&
   $N_L$  &
   $N_R$  &$
   G_L    $&$
   G_R    $&
   models
   \\
\hline
$
  SO(8)  $&$
  SO(8)  $&
  4 &
  4 &$
  SU(7)  $&$
  SU(7)  $&
  \begin{tabular}{c}
  50--53,\\
  55--59
  \end{tabular}
  \\
$
  SO(8)  $&$
  SO(8)  $&
  4 &
  4 &$
  SU(6)\times  SU(2)  $&$
  SU(6)\times  SU(2)  $&
  \begin{tabular}{c}
  44--46,\\
  48,49
  \end{tabular}
  \\
$
  SO(8)  $&$
  SO(8)  $&
  4 &
  4 &$
  SU(5)\times  SU(3)  $&$
  SU(5)\times  SU(3)  $&
  \begin{tabular}{c}
  29--31,\\
  34--36,\\
  39,41
  \end{tabular}
  \\
$
  SO(8)  $&$
  SO(8)  $&
  4 &
  4 &$
  SU(5)\times SU(2)^2  $&$
  SU(5)\times SU(2)^2  $&
  26--28
  \\
$
  SO(8)  $&$
  SO(8)  $&
  4 &
  4 &$
  SU(4)^2  $&$
  SU(4)^2  $&
  22,23
  \\
$
  SO(8)  $&$
  SO(8)  $&
  4 &
  4 &$
  SU(4)\times  SU(3)\times  SU(2)  $&$
  SU(4)\times  SU(3)\times  SU(2)  $&
  16--18
  \\
$
  SO(8)  $&$
  SO(8)  $&
  4 &
  4 &$
  SU(4)\times SU(2)^3  $&$
  SU(4)\times SU(2)^3  $&
  15
  \\
$
  SO(8)  $&$
  SO(8)  $&
  4 &
  4 &$
  SU(3)^3  $&$
  SU(3)^3  $&
  6,7
  \\
$
  SO(8)  $&$
  SO(8)  $&
  4 &
  4 &$
  SU(3)^2\times SU(2)^2  $&$
  SU(3)^2\times SU(2)^2  $&
  2,3
  \\
$
  SO(8)  $&$
  SO(8)  $&
  4 &
  4 &$
  SU(3)\times SU(2)^4  $&$
  SU(3)\times SU(2)^4  $&
  1
  \\
$
  SO(8)  $&$
  SO(4)  $&
  4 &
  2 &$
  SU(3)^3  $&$
  SO(8)\times  SU(2)\times U(1)^3  $&
  10
  \\
$
  SO(6)  $&$
  SO(6)  $&
  0 &
  0 &$
  SU(5)\times U(1)^3  $&$
  SU(5)\times U(1)^3  $&
  37
  \\
$
  SO(6)  $&$
  SO(6)  $&
  0 &
  0 &$
  SU(5)\times U(1)^3  $&$
  SU(3)^2\times U(1)^3  $&
  32,40
  \\
$
  SO(6)  $&$
  SO(6)  $&
  0 &
  0 &$
  SU(4)\times  SU(2)\times U(1)^3  $&$
  SU(4)\times  SU(2)\times U(1)^3  $&
  24
  \\
$
  SO(6)  $&$
  SO(6)  $&
  0 &
  0 &$
  SU(4)\times  SU(2)\times U(1)^3  $&$
  SU(3)\times  SU(2)^2\times U(1)^3  $&
  19
  \\
$
  SO(6)  $&$
  SO(6)  $&
  0 &
  0 &$
  SU(3)^2\times U(1)^3  $&$
  SU(3)^2\times U(1)^3  $&
  8
  \\
$
  SO(6)  $&$
  SO(6)  $&
  0 &
  0 &$
  SU(3)\times SU(2)^2\times U(1)^3  $&$
  SU(3)\times SU(2)^2\times U(1)^3  $&
  4
  \\
$
  SO(6)  $&$
  SO(4)  $&
  0 &
  2 &$
  SU(5)\times U(1)^3  $&$
  SU(3)\times SU(2)^3\times U(1)^3  $&
  54
  \\
$
  SO(6)  $&$
  SO(4)  $&
  0 &
  2 &$
  SU(5)\times U(1)^3  $&$
  SU(3)\times SU(2)\times U(1)^5  $&
  33,42
  \\
$
  SO(6)  $&$
  SO(4)  $&
  0 &
  2 &$
  SU(4)\times  SU(2)\times U(1)^3  $&$
  SU(2)^5\times U(1)^3  $&
  47
  \\
$
  SO(6)  $&$
  SO(4)  $&
  0 &
  2 &$
  SU(4)\times  SU(2)\times U(1)^3  $&$
  SU(2)^3\times U(1)^5  $&
  20,25
  \\
$
  SO(6)  $&$
  SO(4)  $&
  0 &
  2 &$  SU(3)^2\times U(1)^3  $&$
  SU(3)\times  SU(2)\times U(1)^5  $&
  9,11,38
  \\
$
  SO(6)  $&$
  SO(4)  $&
  0 &
  2 &$
  SU(3)\times SU(2)^2\times U(1)^3  $&$
  SU(2)^3\times U(1)^5  $&
  5,21
  \\
$
  SO(6)  $&$
  SO(2)  $&
  0 &
  1 &$
  SU(3)^2\times U(1)^3  $&$
  SU(3)\times U(1)^7  $&
  12
  \\
$
  SO(4)  $&$
  SO(4)  $&
  2 &
  2 &$
  SU(3)\times  SU(2)\times U(1)^5  $&$
  SU(3)\times  SU(2)\times U(1)^5  $&
  13
  \\
$
  SO(4)  $&$
  SO(4)  $&
  0 &
  0 &$
  SU(3)\times U(1)^6  $&$
  SU(3)\times U(1)^6  $&
  43
  \\
$
  SO(4)  $&$
  SO(2)  $&
  2      &
  1      &$
  SU(3)\times  SU(2)\times U(1)^5  $&$
  SU(3)\times U(1)^7  $&
  14
  \\
\hline
\end{tabular}

\end{table}


\end{document}